# Structures of the Reduced Niobium Oxides $Nb_{12}O_{29}$ and $Nb_{22}O_{54}$

T. McQueen[1], Q. Xu[2], E. N. Andersen[1], H.W. Zandbergen[2], and R.J. Cava[1]

[1]Department of Chemistry, Princeton University Princeton NJ 08544
[2]Department of Nanoscience, TU Delft, The Netherlands

## Abstract

The crystal structure of $Nb_{22}O_{54}$ is reported for the first time, and the structure of orthorhombic $Nb_{12}O_{29}$ is reexamined, resolving previous ambiguities. Single crystal x-ray and electron diffraction were employed. These compounds were found to crystallize in the space groups P2/m ($a = 15.7491(2)$ Å, $b = 3.8236(3)$ Å, $c = 17.8521(2)$ Å, $\beta = 102.029(3)°$) and Cmcm ($a = 3.8320(2)$ Å, $b = 20.7400(9)$ Å, $c = 28.8901(13)$ Å) respectively and share a common structural unit, a 4x3 block of corner sharing $NbO_6$ octahedra. Despite different constraints imposed by symmetry these blocks are very similar in both compounds. Within a block, it is found that the niobium atoms are not located in the centers of the oxygen octahedra, but rather are displaced inward toward the center of the block forming an apparent antiferroelectric state. Bond valence sums and bond lengths do not show the presence of charge ordering, suggesting that all 4d electrons are delocalized in these compounds at the temperature studied, T = 200 K.

## Introduction

Niobium and oxygen form a series of complex shear structures between the compositions $Nb_2O_5$ and $Nb_{12}O_{29}$.[1-11] With the exception of $Nb_2O_5$, all contain some fraction of reduced $Nb^{+4}$ ($4d^1$) atoms. Their magnetic and electrical properties have been extensively studied. The electrical conductivity increases as $Nb^{+4}$ content increases, transforming from insulating behavior in $Nb_2O_5$ to metallic behavior in $Nb_{12}O_{29}$.[12-15] At the same time, the oxides become paramagnetic ($Nb_2O_5$ is diamagnetic) and monoclinic $Nb_{12}O_{29}$ even displays an antiferromagnetic transition at $T_N$ = 12 K.[12, 13, 16-19] This makes the shear structures one of the only known series of compounds where $Nb^{4+}$ displays magnetic behavior. The origin of this magnetism, especially the key structural components that give rise to it and the location of the magnetic electron(s), is of interest due to the fact that 4d systems usually form delocalized electronic states rather than localized magnetic ones.

The general structures of the niobium oxides are well-known,[1-4, 6, 20] having been studied with a variety of techniques, including electron, powder neutron, and x-ray diffraction.[21-28] The basic structural unit consists of corner sharing $NbO_6$ octahedra arranged into blocks, as idealized for the 3x3 and 4x3 cases in Figure 1(a). These blocks are then stacked vertically in columns with octahedra from adjacent blocks sharing corners. The columns are then interleaved to form the overall atomic arrangement. Figure 1(b) shows the idealized structure of $Nb_{22}O_{54}$, which is built from 3x3 and 4x3 blocks. Additional Nb atoms lie in tetrahedral sites formed at the intersections of adjacent blocks. Figure 1(c) shows o-$Nb_{12}O_{29}$, which is similar but built solely from 4x3 blocks. The unit cell for each structure is also drawn. Previous work has shown that the Nb atoms are displaced from the ideal positions in the center of the $NbO_6$ octahedra.[2, 4] However, precise single crystal determinations of the displacements have remained elusive due to the difficulty in growing crystals relatively free of stacking faults, point defects and other imperfections. To the authors' knowledge, only the structures of orthorhombic (o-$Nb_{12}O_{29}$) and monoclinic $Nb_{12}O_{29}$ have been previously determined by single crystal work.[2, 3] Here we report the crystal structures of $Nb_{22}O_{54}$ and o-$Nb_{12}O_{29}$ determined by single crystal x-ray diffraction, with the choice of space group for o-$Nb_{12}O_{29}$ confirmed by convergent beam electron diffraction (CBED).

## Experimental

Single crystals of o-$Nb_{12}O_{29}$ and $Nb_{22}O_{54}$ were grown from melts of the corresponding stoichiometric composition with 1-2 mole percent rhodium added to aid crystallization. Starting materials for the syntheses were $Nb_2O_5$ (Aldrich, 99.9%), $NbO_2$ (Johnson-Matthey, 99.8%) and Rh (Aldrich, 99.8%). The starting materials were mixed in appropriate ratios, pressed into pellets, wrapped in Mo foil, and heated in a vacuum furnace back-filled with argon at 1425 °C for 2 hours, followed by cooling at 5 °C/hr to 1300 °C and then a rapid quench to room temperature. EDX analysis was done with a Philips XL-30 SEM equipped with a Princeton Gamma Tech EDX system; rhodium was present in the melts, as expected, but no residual rhodium was observed in the crystals used for this study. Crystals were mounted on a glass fiber with silicone grease, and x-ray data were collected at 200 K with a Nonius KappaCCD diffractometer equipped with a MSC X-stream cryosystem using Mo Kα ($\lambda = 0.71073$Å) radiation and a graphite monochromator. An absorption correction based on the crystal geometry was performed using the numerical Gaussian method.[29] Structures were solved using SHELXS-97 with ab initio direct methods,[30, 31] and refined using SHELXL-97 in the WinGX suite.[32, 33] VaList[34] and PLATON[35, 36] were used for structural analyses. Atomic displacement parameters for all atoms in both structures were refined anisotropically, although minor restraints that did not significantly impact the quality of the fit were used for oxygen atoms at positions O1 and O9-O14 in the final refinement of o-$Nb_{12}O_{29}$. Disorder models for these oxygen atoms did not improve the fit, and were rejected in favor of the thermal parameter restraints. Electron microscopy analysis was performed with a Philips CM300UT electron microscope with a field emission gun, operated at 300 kV, and a link with an EDX element analysis system. Convergent Beam Electron diffraction (CBED) was recorded on image plates with spot sizes less than 10 nm and exposure times ranged from 15-30 s (condenser aperture 10 μm).

**Results**

$Nb_{22}O_{54}$ crystallizes in the monoclinic space group P2/m and has the basic structure shown in Figure 1(b). Refinement statistics for $Nb_{22}O_{54}$ are given in Table 1(a), and complete atomic positions in Table 2. There are no systematic peaks in the residual electron density map and the fit statistics are presumably limited by the presence of stacking faults and other defects. The structure is built from three basic units: Nb atoms in tetrahedral coordination and 3x3 and 4x3 blocks of $NbO_6$ octahedra.

Initial refinements were carried out with the one Nb atom per formula unit in tetrahedral coordination sitting entirely on the Nb12 site, halfway between the niobium planes of adjacent block layers, as shown in Figure 2(a). This site is half filled and there was no evidence of long range order to the filling. However, as shown in the first column of Table 1(b), the R-values for the fit are rather high, as is $\Delta\rho_{\max}$. The residual density for this model revealed a large negative density at the Nb12 position and a large positive density at a neighboring tetrahedral site, labeled Nb13 in Figure 2(a). This indicates that the Nb atom is disordered not only within the Nb12 site, but also across other available tetrahedral positions. The Nb13 site is in the same plane as the niobium atoms that make up the 3x3 blocks. Another tetrahedral site, Nb14, is similarly found in the same plane as the 4x3 blocks. Although there was no observed peak in the residual density at the Nb14 position, refinements were performed to test two models of the disorder of the tetrahedrally-coordinated Nb: (1) across Nb12 and Nb13 and (2) across Nb12, Nb13, and Nb14. The statistics for these refinements are shown in columns two and three of Table 1(b). Disorder across Nb12 and Nb13 results in a substantial drop in the R-values and $\Delta\rho_{max}$, and removes the minima and maxima in the residual density map. Inclusion of disorder onto Nb14 results in only a slight drop in R-values and $\Delta\rho_{max}$. The differences in the atomic positions and thermal parameters of the other atoms between these models and the non-disordered one are within errors and not statistically significant. The Hamilton R-ratio test[37, 38] shows that, at the 95% confidence level, the model that best fits the data includes Nb occupancy on sites Nb12, Nb13, and Nb14. Additionally, the occupancies of the Nb13 and Nb14 sites are both greater than 29 standard deviations away from zero. Thus, this was chosen as the final structural model. The tetrahedrally-coordinated Nb atom is best described as disordered across three different sites. The highest occupied site, Nb12, has 0.806(4) atoms per formula unit (0.403(2) occupancy). The other two sites are Nb13, with 0.135(4) atoms per formula unit (0.068(2) occupancy), and Nb14, with 0.059(2) atoms per formula unit (0.029(1) occupancy). This disorder is consistent with previous high resolution electron microscopy studies in which differences in contrast at the tetrahedral position were observed.[21, 22, 27, 28] All three of these positions are tetrahedrally coordinated to oxygen.

The structure of the 3x3 block of $Nb_{22}O_{54}$ is shown in Figure 2(b). Only five of the niobium atoms are unique (labeled Nb7-Nb11). The other four atoms are related to these by a 2-fold rotation axis in the center of the block. The structure of the 4x3 block is similar, and shown in Figure 3(a). There are six unique niobium atoms (labeled Nb1-Nb6), with the rest related by a 2-fold rotation axis in the center. The atomic displacement parameters of all atoms are within normal limits and do not suggest any systematic errors in the structure.

o-$Nb_{12}O_{29}$ is reported to crystallize in the centered orthorhombic space group Amma,[2] a non-standard setting. For this study, we use the standard setting, Cmcm. Our x-ray diffraction data indicate that the choice of space group for o-$Nb_{12}O_{29}$ is ambiguous. The observed reflections could be indexed with the previously reported cell, but the systematic absences are not consistent with any space group; weak reflections (mean $F^2/\sigma F^2 = 2.70$) were observed at (h0l), l odd, which should be systematically absent in Cmcm. A monoclinic unit cell with $a = 3.8320(2)$ Å, $b = 10.5455(5)$ Å, $c = 28.8901(13)$ Å, and $\gamma = 79.532(2)$ Å, which corresponds to the primitive reduced cell of the orthorhombic choice, also indexed the observed reflections, and the systematic absences were consistent with space group P112/m. Refinements, including twinning in the case of the monoclinic cell, gave equivalently good statistics.

To resolve this ambiguity, CBED was performed, and example patterns are shown in Figure 4. Mirror planes perpendicular to all three axes in the orthorhombic setting are unambiguously observed. Since CBED data comes from a very small region of the specimen, < 10 nm in diameter, it is unlikely that the native symmetry is monoclinic; domains would have to be twinned on this length scale to give the apparent mirror symmetries. As such, the C-centered orthorhombic cell with space group Cmcm was chosen as the correct one.[39] We attribute the weak reflections observed at (h0l), l odd, as originating from monoclinic-$Nb_{12}O_{29}$-like stacking faults, although other possibilities, such as the presence of a small non-merohedral twin, are possible. Despite this, there are no systematic peaks in the residual electron density map, and the fit statistics are reasonable. Refinement statistics for o-$Nb_{12}O_{29}$ are shown in Table 1(a), and complete atomic positions in Table 3. The structure is built entirely from symmetry-equivalent 4x3 blocks of $NbO_6$ octahedra. The arrangement of atoms within a 4x3 block is shown in Figure 3(b). There are six unique niobium atoms (labeled Nb1-Nb6), with the rest related by a mirror plane through the center of the block.

**Discussion**

$Nb_{22}O_{54}$ and o-$Nb_{12}O_{29}$ share a common motif, a 4x3 block of corner-sharing $NbO_6$ octahedra. Despite the different symmetry constraints on both the Nb and O atoms, the observed structures of these blocks are remarkably similar. The Nb atoms are displaced from the centers of the oxygen octahedra, towards the center of the blocks (see Figure 3 top views). These Nb displacements are characteristic of spontaneous polarization and the formation of electric dipoles, as in many $Nb^{5+}$ ($4d^0$) ferroelectrics. However, in $Nb_{22}O_{54}$ and o-$Nb_{12}O_{29}$ the displacements are arranged so that individual blocks, and hence the material, have no net electric polarization; thus the two compounds are antiferroelectric. The 3x3 blocks, present only in $Nb_{22}O_{54}$, also show distortions toward the center of the blocks, resulting in similar behavior. Although some of the resultant Nb-O bonds are relatively short (1.76-1.85 Å), they are no shorter than bond lengths observed in other $Nb^{5+}$ oxides, and longer than the ones reported for monoclinic $Nb_{12}O_{19}$.[15, 17, 40-42]

Bond valence sums, listed in Table 4, were used to try to determine whether there is a specific location for the reduced Nb in $Nb_{22}O_{54}$ and o-$Nb_{12}O_{29}$. However, all bond valence sums are very close to 5+ with no one site significantly different from the others in either compound. Although the Nb13 and Nb14 tetrahedral sites in $Nb_{22}O_{54}$ have lower bond valences, their small fractional occupancies mean that the average valence of the tetrahedrally coordinated Nb atom is still very close to 5+. Additionally, the observed magnetic moment corresponds to one localized electron ($Nb^{4+}$) per $Nb_{22}O_{54}$ unit, but the occupancy of the Nb14 site would correspond to only 0.059(2) localized electrons per $Nb_{22}O_{54}$ unit. This lack of a distinct $Nb^{4+}$ site sufficient to explain the observed magnetic moment suggests that the valence electrons responsible for the magnetism are not localized on individual Nb atoms, at least at T = 200 K. This is consistent with a previous report on the monoclinic form of $Nb_{12}O_{29}$, which showed no charge ordering at ambient temperature, but does not explain the observed magnetic behavior of these materials.[13]

## Conclusions

The structures of $Nb_{22}O_{54}$ and o-$Nb_{12}O_{29}$ were solved using single crystal x-ray diffraction. No localization of $Nb^{4+}$ atoms, based on bond valence sums, was observed for either material. The 3x3 and 4x3 blocks of corner sharing $NbO_6$ octahedra have clear displacements of the Nb atoms in towards the center of the blocks, forming an antiferroelectric ordering of the electric dipoles. The structures of the 4x3 blocks in each compound are similar, suggesting a particular stability of the block motif. We speculate that each magnetic electron may be delocalized across a 4x3 block, as that would explain the inability to locate the reduced site with bond valence sums. Theoretical calculations on the 4x3 blocks would be of interest to understand why the Nb atoms displace toward the center of the blocks in an antiferroelectric fashion, and to explain the observed magnetic behavior in these materials.

## Acknowledgements

T. McQueen gratefully acknowledges support by the national science foundation graduate research fellowship program. This work was done under NSF grant DMR06-20234.

**Table 1**. (a) Final crystal, data collection, and refinement statistics for $Nb_{22}O_{54}$ and o-$Nb_{12}O_{29}$. (b) Comparison of different Nb atom disorder model refinements of $Nb_{22}O_{54}$.

| **(a) Final Refinements** | $Nb_{22}O_{54}$ | o-$Nb_{12}O_{29}$ |
|---:|:---:|:---:|
| Space group | P 1 2/m 1 | C m c m |
| *a* (Å) | 15.7491(2) | 3.8320(2) |
| *b* (Å) | 3.8236(3) | 20.7400(9) |
| *c* (Å) | 17.8521(2) | 28.8901(13) |
| β (°) | 102.029(3) | - |
| V (Å$^3$) | 1051.42(8) | 2296.06(19) |
| Z | 1 | 4 |
| μ (mm$^{-1}$) | 5.88 | 5.87 |
| Crystal (mm$^3$) | 0.030 x 0.034 x 0.100 | 0.020 x 0.030 x 0.055 |
| $T_{min}$ / $T_{max}$ | 0.7387 / 0.8592 | 0.7590 / 0.8852 |
| Measured Reflections | 14057 | 11206 |
| Independent Reflections | 3504 | 1944 |
| Reflections ($F^2 > 2\sigma F^2$) | 2712 | 1394 |
| R ($F^2 > 2\sigma F^2$) | 0.0595 | 0.0744 |
| R (all) | 0.0829 | 0.0941 |
| wR2($F^2$) | 0.1269 | 0.1922 |
| GooF | 1.190 | 1.200 |
| $\Delta\rho_{max}/\Delta\rho_{rms}$ (e/Å$^3$) | 4.803 / 0.376 | 3.41 / 0.575 |
| wR2 weighting | $\frac{1}{\sigma^2(F_0^2)+(0.0265P)^2+4.44P}$ | $\frac{1}{\sigma^2(F_0^2)+(0.000P)^2+397.87P}$ |
| Parameters | 241 | 127 |

| **(b) $Nb_{22}O_{54}$** | Nb12 Only | Nb12, Nb13 | Nb12, Nb13, Nb14 |
|---:|:---:|:---:|:---:|
| R ($F^2 > 2\sigma F^2$) | 0.0713 | 0.0612 | 0.0595 |
| R (all) | 0.0943 | 0.0844 | 0.0829 |
| wR2($F^2$) | 0.1699 | 0.1346 | 0.1269 |
| GooF | 1.160 | 1.182 | 1.190 |
| $\Delta\rho_{max}/\Delta\rho_{rms}$ (e/Å$^3$) | 10.44 / 0.472 | 4.88 / 0.384 | 4.803 / 0.376 |
| Parameters | 234 | 238 | 241 |

**Table 2**. Structural parameters for $Nb_{22}O_{54}$ at 200 K. Bond lengths are shown in Figures 2 and 3. Anisotropic atomic displacement parameters are also given. For all atoms, $U_{12} = U_{23} = 0$.

| Label | Site | x | y | z | Occ. | $U_{11}$ | $U_{22}$ | $U_{33}$ | $U_{13}$ |
|---|---|---|---|---|---|---|---|---|---|
| **Nb1** | 2m | 0.54947(4) | 0 | 0.42869(4) | 1 | 0.0111(3) | 0.0089(4) | 0.0084(3) | 0.0012(2) |
| **Nb2** | 2m | 0.69313(4) | 0 | 0.63439(4) | 1 | 0.0115(3) | 0.0108(4) | 0.0098(4) | 0.0023(3) |
| **Nb3** | 2m | 0.16133(4) | 0 | 0.16546(4) | 1 | 0.0128(3) | 0.0116(4) | 0.0119(4) | 0.0035(3) |
| **Nb4** | 2m | 0.04207(4) | 0 | 0.74397(4) | 1 | 0.0109(3) | 0.0091(4) | 0.0080(3) | 0.0009(3) |
| **Nb5** | 2m | 0.89929(5) | 0 | 0.54279(4) | 1 | 0.0107(3) | 0.0406(6) | 0.0077(4) | 0.0007(3) |
| **Nb6** | 2m | 0.75733(4) | 0 | 0.34239(4) | 1 | 0.0113(3) | 0.0089(4) | 0.0083(4) | 0.0003(3) |
| **Nb7** | 2n | 0.44538(4) | 0.5 | 0.28596(4) | 1 | 0.0112(3) | 0.0091(4) | 0.0079(3) | 0.0009(3) |
| **Nb8** | 2n | 0.64906(4) | 0.5 | 0.19901(4) | 1 | 0.0119(3) | 0.0091(4) | 0.0075(4) | -0.0007(3) |
| **Nb9** | 2n | 0.15002(4) | 0.5 | 0.8877(4) | 1 | 0.0121(3) | 0.0101(4) | 0.0087(4) | 0.0007(3) |
| **Nb10** | 2n | 0.29726(4) | 0.5 | 0.08644(4) | 1 | 0.0102(3) | 0.0100(4) | 0.0093(4) | 0.0019(3) |
| **Nb11** | 2n | 0.5 | 0.5 | 0 | 1 | 0.0112(5) | 0.0364(8) | 0.0078(5) | 0.0006(4) |
| **Nb12** | 2i | 0 | 0.2198(6) | 0 | 0.403(2) | 0.0094(8) | 0.0403(14) | 0.0082(9) | 0.0010(6) |
| **Nb13** | 2n | 0.0617(7) | 0.5 | 0.0294(6) | 0.0677(18) | 0.0094(8) | 0.0403(14) | 0.0082(9) | 0.0010(6) |
| **Nb14** | 2m | 0.9807(16) | 0 | 0.0450(15) | 0.0295(18) | 0.0094(8) | 0.0403(14) | 0.0082(9) | 0.0010(6) |
| **O1** | 2m | 0.4452(3) | 0 | 0.3213(3) | 1 | 0.013(3) | 0.006(3) | 0.009(3) | -0.001(2) |
| **O2** | 2n | 0.5209(3) | 0.5 | 0.4068(3) | 1 | 0.011(2) | 0.010(3) | 0.009(3) | -0.001(2) |
| **O3** | 2m | 0.5848(3) | 0 | 0.5349(3) | 1 | 0.018(3) | 0.016(3) | 0.008(3) | 0.002(2) |
| **O4** | 2m | 0.6463(3) | 0 | 0.3920(3) | 1 | 0.016(3) | 0.012(3) | 0.015(3) | 0.004(2) |
| **O5** | 2n | 0.6583(3) | 0.5 | 0.6444(3) | 1 | 0.007(2) | 0.012(3) | 0.015(3) | 0.005(2) |
| **O6** | 2m | 0.7540(3) | 0 | 0.7357(3) | 1 | 0.017(3) | 0.012(3) | 0.012(3) | 0.000(2) |
| **O7** | 2m | 0.7778(4) | 0 | 0.5838(3) | 1 | 0.016(3) | 0.023(4) | 0.012(3) | 0.004(2) |
| **O8** | 2n | 0.1869(4) | 0.5 | 0.1494(3) | 1 | 0.017(3) | 0.011(3) | 0.017(3) | 0.010(2) |
| **O9** | 2m | 0.2749(4) | 0 | 0.1033(3) | 1 | 0.015(3) | 0.013(3) | 0.018(3) | 0.004(2) |
| **O10** | 2m | 0.0960(4) | 0 | 0.0566(3) | 1 | 0.025(3) | 0.019(4) | 0.014(3) | 0.003(2) |
| **O11** | 2m | 0.0659(3) | 0 | 0.2043(3) | 1 | 0.015(3) | 0.011(3) | 0.017(3) | 0.008(2) |
| **O12** | 2n | 0.0568(3) | 0.5 | 0.7749(3) | 1 | 0.011(3) | 0.014(3) | 0.008(3) | -0.002(2) |
| **O13** | 2m | 0.1226(4) | 0 | 0.8649(3) | 1 | 0.015(3) | 0.008(3) | 0.017(3) | -0.002(2) |
| **O14** | 2m | 0.1465(3) | 0 | 0.7073(3) | 1 | 0.012(3) | 0.013(3) | 0.014(3) | 0.004(2) |
| **O15** | 2m | 0.9691(3) | 0 | 0.6513(3) | 1 | 0.012(3) | 0.019(4) | 0.012(3) | 0.000(2) |
| **O16** | 2n | 0.8877(4) | 0.5 | 0.5502(3) | 1 | 0.017(3) | 0.016(4) | 0.023(4) | 0.004(3) |
| **O17** | 2m | 1 | 0 | 0.5 | 1 | 0.015(4) | 0.030(6) | 0.008(4) | 0.004(3) |
| **O18** | 2m | 0.8227(3) | 0 | 0.4388(3) | 1 | 0.013(3) | 0.014(3) | 0.011(3) | 0.000(2) |
| **O19** | 2n | 0.7315(3) | 0.5 | 0.3189(3) | 1 | 0.014(3) | 0.010(3) | 0.010(3) | 0.000(2) |
| **O20** | 2m | 0.6649(3) | 0 | 0.2288(3) | 1 | 0.014(3) | 0.007(3) | 0.010(3) | -0.002(2) |
| **O21** | 2n | 0.9616(4) | 0.5 | 0.0670(4) | 1 | 0.016(3) | 0.035(4) | 0.018(3) | 0.005(2) |
| **O22** | 2n | 0.2155(4) | 0.5 | 0.9870(3) | 1 | 0.016(3) | 0.015(3) | 0.011(3) | -0.002(2) |
| **O23** | 2n | 0.2445(3) | 0.5 | 0.8402(3) | 1 | 0.015(3) | 0.015(3) | 0.013(3) | 0.002(2) |
| **O24** | 2n | 0.3687(3) | 0.5 | 0.1960(3) | 1 | 0.015(3) | 0.012(3) | 0.013(3) | 0.003(2) |
| **O25** | 2n | 0.3930(3) | 0.5 | 0.0461(3) | 1 | 0.015(3) | 0.019(4) | 0.010(3) | 0.004(2) |
| **O26** | 2n | 0.5460(3) | 0.5 | 0.2536(3) | 1 | 0.014(3) | 0.011(3) | 0.014(3) | 0.006(2) |
| **O27** | 2n | 0.5773(3) | 0.5 | 0.1063(3) | 1 | 0.014(3) | 0.008(3) | 0.014(3) | 0.003(2) |
| **O28** | 2m | 0.5 | 0 | 0 | 1 | 0.018(4) | 0.015(5) | 0.023(5) | 0.003(4) |

**Table 3**. Structural parameters for o-$Nb_{12}O_{29}$ at 200 K. All sites are fully occupied. Bond lengths are shown in Figure 3. Anisotropic atomic displacement parameters are also given. For all atoms, $U_{13} = U_{23} = 0$.

| Label | Site | x | y | z | $U_{11}$ | $U_{22}$ | $U_{33}$ | $U_{12}$ |
|---|---|---|---|---|---|---|---|---|
| **Nb1** | 8f | 0 | 0.46238(7) | 0.54992(5) | 0.0070(7) | 0.0066(6) | 0.0061(6) | 0.0002(5) |
| **Nb2** | 8f | 0 | 0.64757(7) | 0.54832(5) | 0.0067(7) | 0.0059(6) | 0.0054(6) | -0.0002(5) |
| **Nb3** | 8f | 0 | 0.83115(7) | 0.54939(5) | 0.0040(7) | 0.0057(6) | 0.0058(6) | 0.0007(5) |
| **Nb4** | 8f | 0 | 0.83160(7) | 0.68400(5) | 0.0019(6) | 0.0032(6) | 0.0075(6) | 0.0002(5) |
| **Nb5** | 8f | 0 | 0.64779(8) | 0.68440(5) | 0.0391(11) | 0.0056(6) | 0.0061(6) | 0.0006(6) |
| **Nb6** | 8f | 0 | 0.46416(7) | 0.68470(5) | 0.0067(7) | 0.0054(6) | 0.0063(6) | 0.0000(5) |
| **O1** | 8f | 0 | 0.3562(6) | 0.5354(4) | 0.005(3) | 0.005(3) | 0.007(3) | 0.0008(18) |
| **O2** | 8f | 0 | 0.9385(6) | 0.5431(4) | 0.009(6) | 0.005(5) | 0.017(6) | 0.003(4) |
| **O3** | 8f | 0 | 0.5491(6) | 0.5298(4) | 0.007(6) | 0.009(5) | 0.011(5) | 0.000(4) |
| **O4** | 8f | 0 | 0.4655(6) | 0.6115(4) | 0.010(6) | 0.011(5) | 0.007(5) | 0.003(4) |
| **O5** | 8f | 0 | 0.1502(6) | 0.5284(4) | 0.008(6) | 0.011(5) | 0.006(5) | 0.001(4) |
| **O6** | 8f | 0 | 0.7436(6) | 0.5414(4) | 0.012(7) | 0.008(5) | 0.012(6) | -0.001(4) |
| **O7** | 8f | 0 | 0.6418(6) | 0.6091(4) | 0.014(6) | 0.006(5) | 0.013(5) | 0.004(5) |
| **O8** | 8f | 0 | 0.8373(6) | 0.6113(4) | 0.011(6) | 0.012(5) | 0.008(5) | 0.001(4) |
| **O9** | 8f | 0 | 0.3576(6) | 0.6791(4) | 0.006(3) | 0.006(3) | 0.007(3) | 0.0002(18) |
| **O10** | 8f | 0 | 0.9403(5) | 0.6799(4) | 0.003(3) | 0.004(3) | 0.005(3) | 0.0011(18) |
| **O11** | 4c | 0 | 0.8347(7) | 0.75 | 0.003(3) | 0.004(3) | 0.004(3) | 0 |
| **O12** | 8f | 0 | 0.7434(7) | 0.6786(5) | 0.019(3) | 0.017(3) | 0.018(3) | 0.0002(19) |
| **O13** | 8f | 0 | 0.1488(7) | 0.6758(4) | 0.013(3) | 0.015(3) | 0.015(3) | -0.0004(19) |
| **O14** | 4c | 0 | 0.6503(9) | 0.75 | 0.011(4) | 0.010(4) | 0.010(4) | 0 |
| **O15** | 8f | 0 | 0.5512(6) | 0.6838(4) | 0.007(6) | 0.008(5) | 0.013(5) | 0.000(4) |
| **O16** | 4c | 0 | 0.4568(9) | 0.75 | 0.026(12) | 0.011(8) | 0.005(7) | 0 |

**Table 4**. Bond valence parameters for both compounds, calculated assuming a 5+ valence for Nb. The sites Nb1 through Nb6 correspond to the 4x3 block (Figure 3) and Nb7 through Nb11 correspond to the 3x3 block (Figure 2(b)). Nb12 through Nb14 together represent the one atom per formula unit in $Nb_{22}O_{54}$ that is tetrahedrally coordinated. The average bond valence for this Nb atom, calculated using the relative occupancies of the three sites, is also listed.

| **Site** | **$Nb_{22}O_{54}$** | **o-$Nb_{12}O_{29}$** |
|---|---|---|
| Nb1 | 4.93 | 4.85 |
| Nb2 | 4.88 | 4.70 |
| Nb3 | 5.02 | 4.87 |
| Nb4 | 4.98 | 4.81 |
| Nb5 | 4.97 | 5.00 |
| Nb6 | 4.95 | 5.04 |
| | | |
| Nb7 | 4.98 | |
| Nb8 | 4.95 | |
| Nb9 | 5.01 | |
| Nb10 | 4.98 | |
| Nb11 | 4.92 | |
| | | |
| Nb12 | 5.13 | |
| Nb13 | 4.57 | |
| Nb14 | 4.13 | |
| average Nb12-14 | 5.00 | |

**Figure 1.** (a) The idealized building blocks for $Nb_{22}O_{54}$ and o-$Nb_{12}O_{29}$, as viewed from the top. (b) The structure of $Nb_{22}O_{54}$ built from the basic units shown in (a), with extra Nb atoms in the tetrahedral holes between blocks. The unit cell is drawn. (c) The structure of o-$Nb_{12}O_{29}$, along with the expected unit cell. In both (b) and (c), the lighter blocks are displaced by ½ the unit cell along the viewing direction.

**Figure 2.** Actual structures of the (a) tetrahedral and (b) 3x3 block portions of $Nb_{22}O_{54}$. All thermal ellipsoids are drawn at the 50% probability level. The smaller graphs in the upper area of (a) and (b) show the units as viewed from the same perspective as Figure 1. The niobium atoms at the center of the 3x3 block are displaced from the middle of the oxygen octahedral toward the center of the block.

**Figure 3.** Actual structures of the 4x3 block portions of (a) $Nb_{22}O_{54}$ and (b) o-$Nb_{12}O_{29}$. All thermal ellipsoids are drawn at the 50% probability level. The smaller graphs in the upper area of (a) and (b) show the units as viewed from the same perspective as Figure 1. The niobium atoms are displaced from the middle of the oxygen octahedra toward the center of the block, forming an antiferroelectric state. Despite the large differences between the two compounds, these blocks have an almost identical structure.

**Figure 4.** CBED images of o-$Nb_{12}O_{29}$, taken along two crystallographic axes as well as in-between, showing mirror planes perpendicular to all three principle axes. The size of the illuminated area is < 10 nm. This data shows that the Cmcm orthorhombic unit cell is the correct one for o-$Nb_{12}O_{29}$, and indexing is given according to that cell. The contrast of the CBED was adjusted to show both low order and high order reflections.[43]

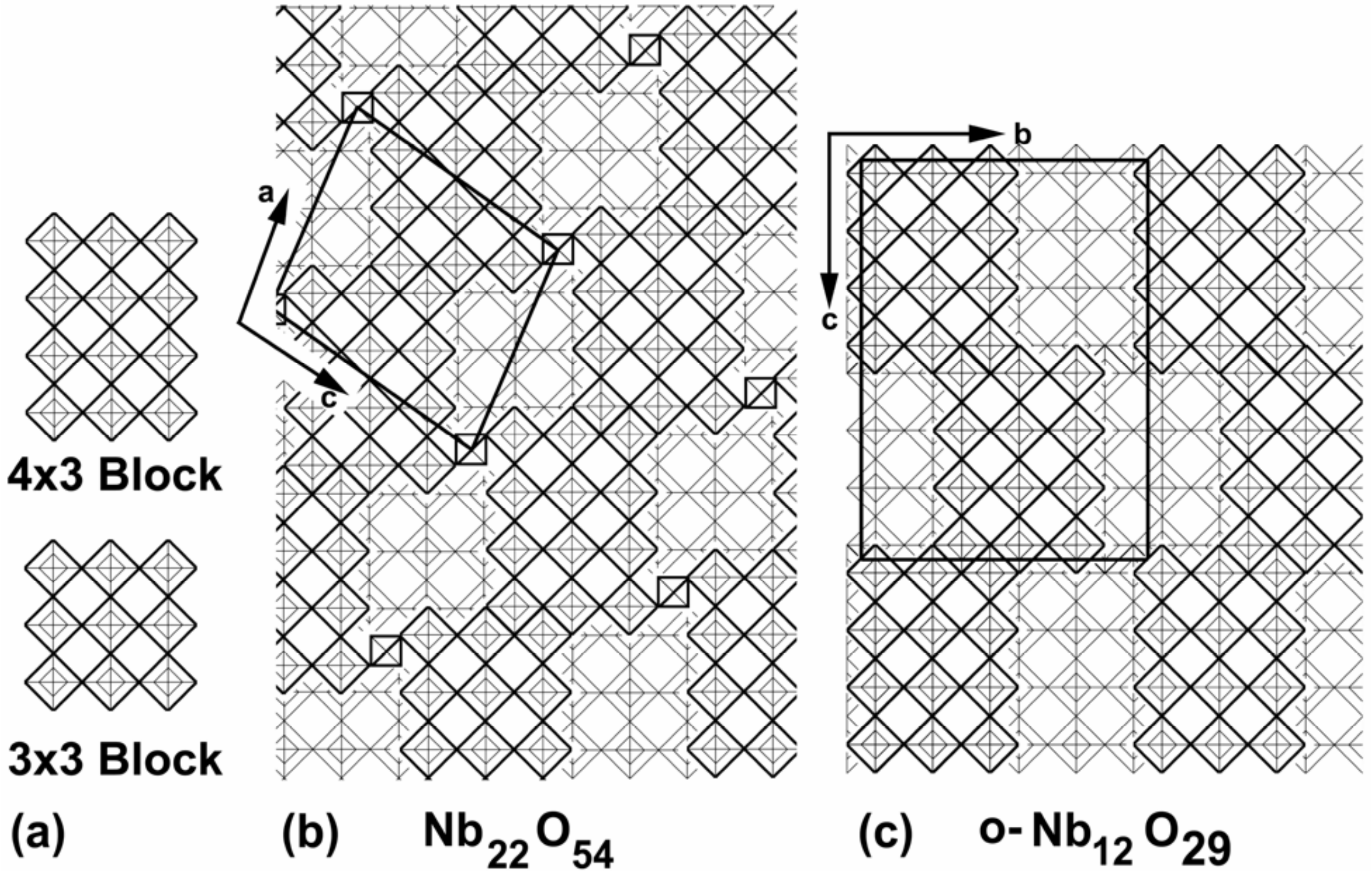

Figure 1

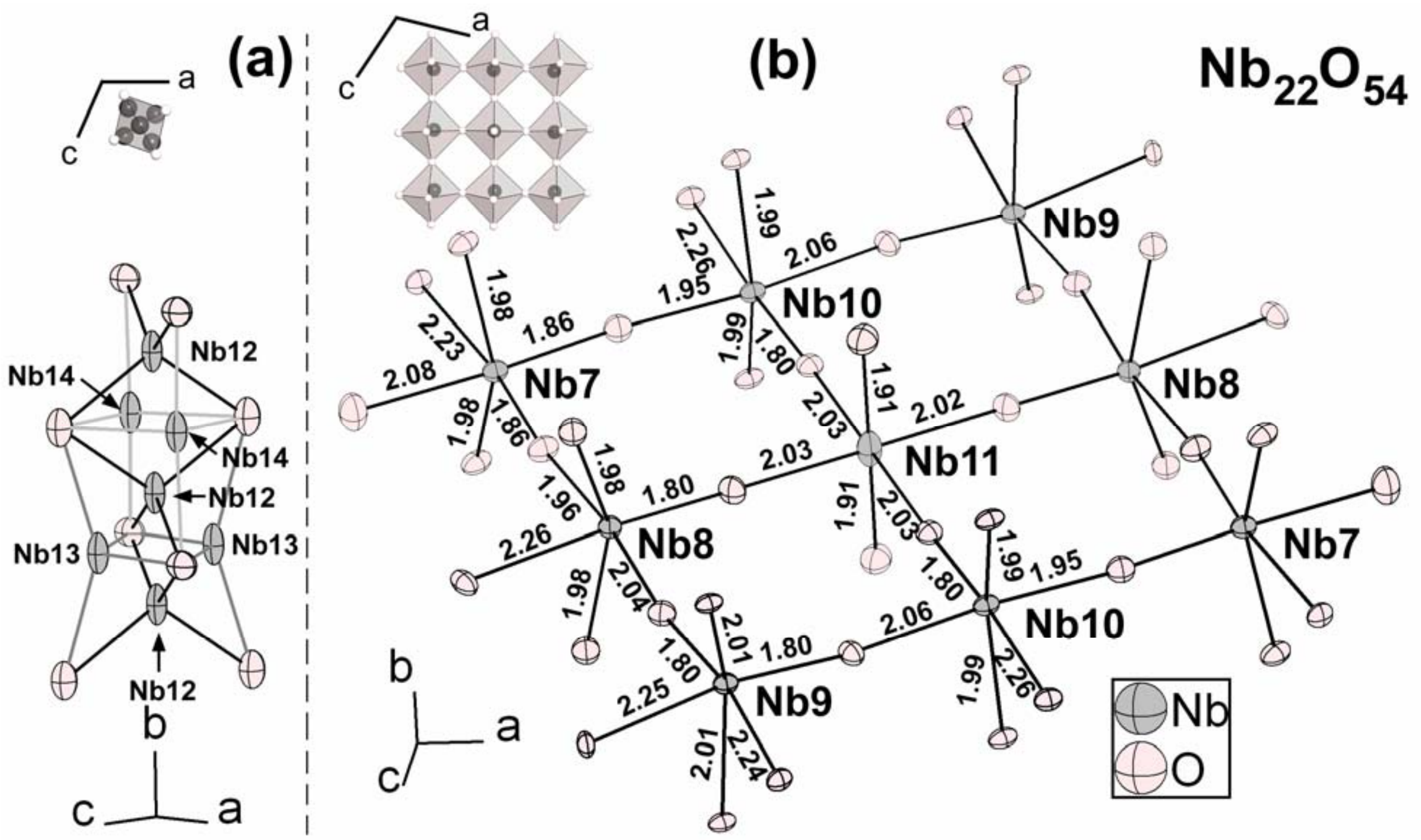

Figure 2

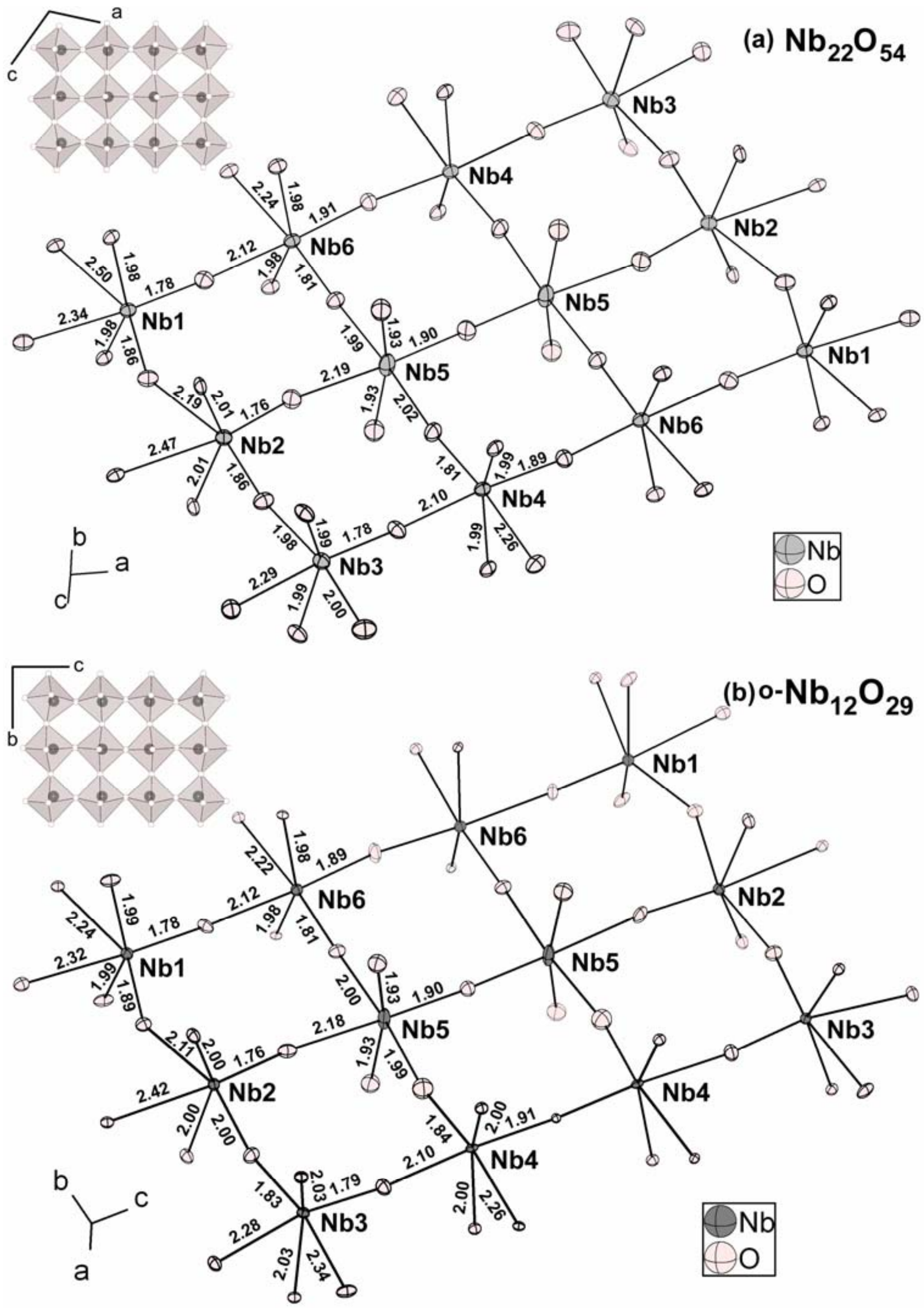

Figure 3

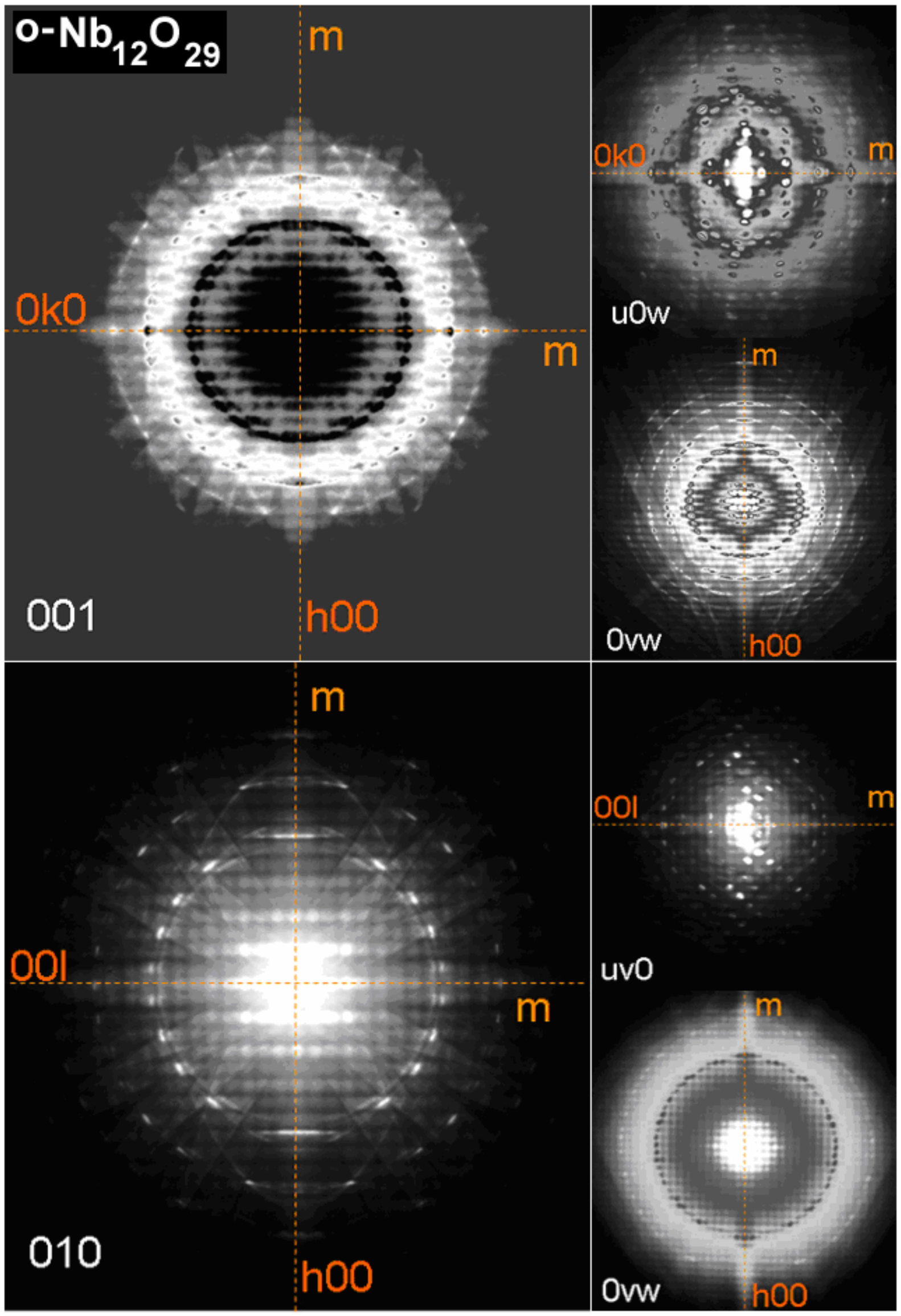
Figure 4